\documentclass[twocolumn,aps,PRE]{revtex4}
\usepackage{graphicx}
\usepackage{amsmath}
\usepackage{todo}
\usepackage{nicefrac}

\usepackage{color}

\newcommand{\myexp}{\mathrm{e}}

\newcommand{\tobs}{t_{\mathrm{obs}}}
\newcommand{\Boltz}{k_{\mathrm{B}}}
\newcommand{\tauB}{\tau_{\mathrm{B}}}

\makeatletter
\g@addto@macro\@floatboxreset\centering
\makeatother

\newcommand{\Jbar}[1]{\overline{#1}}

\begin{document}
\author{Ian R. Thompson}
\author{Robert L. Jack}
\address{Department of Physics, University of Bath, Bath, BA2 7AY, United Kingdom}
\title{Dynamical phase transitions in one-dimensional hard-particle systems}

\begin{abstract}
We analyse a one-dimensional model of hard particles, within ensembles of trajectories that are conditioned (or biased) to atypical values of the time-averaged dynamical activity.  We analyse two phenomena that are associated with these large deviations of the activity: phase separation (at low activity) and the formation of hyperuniform states (at high activity).  We consider a version of the model which operates at constant volume, and a version at constant pressure.  In these non-equilibrium systems, differences arise between the two ensembles, because of the extra freedom available to the constant-pressure system, which can change its total density.  We discuss the relationships between different ensembles, mechanical equilibrium, and the probability cost of rare density fluctuations.%

\end{abstract}

\maketitle

\section{Introduction}

Out-of-equilibrium systems differ from their equilibrium counterparts in many ways.  
For example, long-ranged correlations are generic in non-equilibrium steady states~\cite{Spohn:1983dt,Derrida:2007hr}; unusual phase transitions can take place~\cite{Hinrichsen:2000cy}; and there may be significant differences in behavior for the same system in different ensembles (for example, canonical and grand canonical~\cite{Cohen:2012hx,Cohen:2014je}).
Recently, there has been considerable interest in large-deviation phenomena~\cite{Touchette:2009lq}, based on ensembles of trajectories that are conditioned on atypical values of time-averaged observables~\cite{Lecomte:2005eq,Tailleur:2007by,Hedges:2009hb,Speck:2012df}. 
In glassy systems, ensembles conditioned on the dynamical activity quite generically support phase transitions between active and inactive states~\cite{Garrahan:2007gw,Hedges:2009hb}.  
In other systems, such biases can result in phase separation~\cite{Lecomte:2012en}, and other kinds of phase transition~\cite{Lecomte:2005eq}.  

Here, we consider large deviations of the activity in a model of hard particles that diffuse in one dimension.  
Considering this system in a constant-volume ensemble, we showed previously~\cite{Jack:2015cx} that it supports both phase-separated and hyperuniform states (in which large-scale density fluctuations are strongly suppressed~\cite{Torquato:2003fk}).  
In contrast to equilibrium systems, these biased ensembles of trajectories support coexistence of phases with different (mechanical) pressures: similar effects have recently been discussed in active matter systems~\cite{Wittkowski:2014dt,solon-press}. 
We analyze this effect using a Langevin equation for the time-evolution of the system -- we explain it in terms of random forces that acquire non-zero averages in conditioned ensembles of trajectories.  

We also consider a version of the system in the constant-pressure ensemble.  
At equilibrium, the behavior in such a system resembles that of a subsystem of a large constant-volume system, and phase separation in constant-density
systems is accompanised by bistable behaviour in the constant-pressure ensemble.  
However, in the non-equilibrium ensembles considered here, this familiar picture breaks down.  We interpret this effect in terms of phase coexistence
between states with
different pressures.

Taken together, our results highlight the broad range of phenomena that can occur in conditioned ensembles of trajectories, even in simple models.  
They show that equilibrium ideas of pressure and ensemble-equivalence can sometimes be applied in these contexts, but that these applications may be subtle and require careful justification.
In the following, Section \ref{sec:Models} describes the models and methods that we will use, Section \ref{sec:NVT_results} shows results for constant-volume systems, while Section \ref{sec:NPT_results} includes results at constant-pressure. In Section~\ref{sec:virial}, we discuss force balance and mechanical equilibrium
in these systems, based around the virial pressure.
We draw our conclusions in Section \ref{sec:conclusion}.  

\section{Model and Methods}\label{sec:Models}

\subsection{Model}\label{subsec:Models}

We consider a system of $N$ hard particles undergoing Brownian motion in one dimension, with periodic boundaries. 
Particle motion is described by the overdamped Langevin equation:
\begin{align}
\dot{x_i}=-\frac{D_{p}}{k_{B}T}\nabla_i U +\sqrtsign{2D_p} \eta_i(t) \label{particle_langevin}
\end{align}
where $x_i$ is the position of particle $i$, $U$ is the potential energy of the system, and $D_p$ is the diffusion coefficient for particle motion. 
The random noise $\eta_i$ has zero mean and is uncorrelated in time and space:
\begin{align}
\left<\eta_{i}(t)\eta_{j}(t')\right>=\delta(t-t')\delta_{ij}  .
\end{align}

We consider the case where the energy $U$ is given by a pair potential, $U=\frac12 \sum_{i\neq j} v(x_{i} - x_{j})$.  Specifically, we consider hard particles of size $l_0$,
\begin{align}
v(x)=\begin{cases}
    \infty,& \text{if } x\leq l_0\\
    0,              & \text{otherwise}.
\end{cases}
\label{equ:vx}
\end{align}
In this case, particles cannot interpenetrate or otherwise move past one another.  
[For the derivatives in (\ref{particle_langevin}) to make sense, one should regularize the potential by smoothing its discontinuities, so that the hard-particle case may be obtained by taking a suitable limit. 
In practice, we will simulate the time evolution of (\ref{particle_langevin}) using a Monte Carlo scheme, so no explicit regularisation is required.]

In one dimension, the exact equation of state for this system is that of an ideal gas with excluded volume:
\begin{equation}
 P(L-Nl_0) =N\Boltz T 
\end{equation}
where $P$ is the pressure (we fix the units of energy by setting $k_{\rm B}T=1$ so that
the units of pressure are $l_0^{-1}$).  We define the packing fraction $\phi =\frac{Nl_0}{L}$. 

We simulate this system in both the $NVT$ ensemble (constant-density), and the $NPT$ ensemble (constant-pressure).  For simulations at constant-pressure $P$,
the system size $L$ evolves as
\begin{align}
\dot{L}=-\frac{ D_{L}}{k_{B}T}[P+(\partial U/\partial L)]+\sqrtsign{2D_{L}}\eta_L(t) 
 \label{eqn:volume_langevin}
\end{align}
where 
$D_{L}$ is the diffusion coefficient for the volume coordinate. 
The noise term $\eta_{L}$ has zero mean, with $\left<{\eta_L}(t){\eta_L}(t')\right>=\delta(t-t')$, and $\eta_L(t)$ is independent of the other noises ${\eta_i}(t)$.

To determine an appropriate value for $D_L$, we use a hydrodynamic argument.
For large systems, the system size is almost always close to its mean value $\Jbar{L} = \langle L \rangle$, with small fluctuations of size $O(\sqrt{N})$.
Noting that $(\partial U/\partial L)=L^{-1} \sum_i (r_{i+1}-r_{i})v'(r_{i+1}-r_i)$ is the negative of the virial pressure, 
one may therefore linearise (\ref{eqn:volume_langevin}), taking $[P+(\partial U/\partial L)] \approx \frac{1}{\kappa_T \Jbar{L}}(L-\Jbar{L})$ where
$\kappa_T = (-1/\Jbar{L})(\partial L/\partial P)$ is the isothermal compressibility.  From the linearised Langevin equation, the volume relaxation time
is $\tau_{LL} = (\Jbar{L}\kappa_T k_{B}T/D_{L})$.   Diffusive scaling indicates that this relaxation time should be
of order $\Jbar{L}^2/((2\pi)^2 D_c)$ where $D_c$ is a collective diffusion constant.  
Setting $\tau_{LL} = \Jbar{L}^2/((2\pi)^2 D_c)$ and assuming $D_c\simeq D_p$ yields
\begin{equation}
D_L \simeq (2\pi)^2 \frac{D_p \kappa_T k_B T }{\Jbar{L}}  \label{eqn:diffusion_relation}
\end{equation}
The numerical prefactor in this equation is not crucial for this work but the scaling of $D_L$ with system size $\Jbar{L}$ will be important in what follows.

\subsection{Monte Carlo dynamics}\label{subsec:MC_dynamics}

We use a Monte Carlo (MC) method to implement these dynamical Langevin equations.
The system evolves by single particle MC moves~\cite{Frenkel:2002book}.  
Particle displacements are chosen uniformly from the range $-S\leq\Delta x\leq S$, with $S=0.1l_0$.
If the displacement results in a particle overlap, it is rejected; otherwise, it is accepted.

The natural unit of time in the system is $\tauB$, the time taken for the root mean squared displacement of a free particle to reach its own length: 
\begin{align}
\tauB= \frac{l_0^{2}}{2D_p}.
\end{align}
In terms of the Monte Carlo dynamics, this time interval corresponds to 
\begin{align}
N_{\mathrm{MC}}=\frac{N l_{0}^{2}}{S^{2}/3}
\end{align}
attempted MC moves (here $S^2/3$ is the mean squared displacement for a single step).  This choice ensures that free-particle diffusion always has diffusion constant $D_p$, as required.  The Monte Carlo
dynamical scheme is equivalent to solving the Langevin equation (\ref{particle_langevin}), in the limit where the step size $S\to 0$ \cite{Berthier:2007jj}. The value of $S$
used here is small enough that qualitative features do not depend on $S$.

When the packing fraction $\phi$ is large, the rate of acceptance of MC moves can get small, and it becomes convenient to use a rejection free MC algorithm that operates in continuous time\cite{Bortz:1975kc,Frenkel:2002book}.
To achieve this,
all possible particle displacements are calculated: let $g_i$ denote the fraction of possible moves
for particle $i$ that are compatible with the hard-particle interactions (i.e., the fraction that would be accepted). 
Particle $i$ is selected with probability $g_i/(\sum_j g_j)$ and one of its possible moves is implemented.
The simulation time is then incremented by
\begin{align}
	t_{i}=\frac{\tau_{\rm B}}{N_{\rm MC}} \cdot \frac{N}{\sum_j g_j}  \cdot  \ln(1/\mu) 
\end{align}
 where $\mu$ is randomly distributed $0<\mu\leq 1$ (so $\ln(1/\mu) $ is exponentially distributed with a mean of unity). 
The process is repeated and moves are made until the total simulation time reaches the desired duration.

In constant-pressure simulations, we also perform MC moves in which we propose changes to the volume of the system. 
A change in volume is proposed as $L_{\mathrm{new}}=L_{\mathrm{old}}+\Delta L$ where $-S_{L}\leq\Delta L\leq S_{L}$ and the positions of particles are scaled uniformly by $\frac{L_{\mathrm{new}}}{L_{\mathrm{old}}}$. 
If this causes any particles to overlap the move is rejected immediately.
Otherwise, the move is accepted with probability $\min(1,e^{-\beta P\Delta L+N\ln(L_{\mathrm{new}}/L_{\mathrm{old}}}))$~\cite{Frenkel:2002book}.

At every MC step, a volume change is proposed with fixed probability $1/N$ and a particle displacement is proposed with probability $(N-1)/N$.
To satisfy equation (\ref{eqn:diffusion_relation}) we specify that the maximum volume change satifisfies $S_{L}^{2}=S^{2}(2\pi)^{2}\Boltz T\kappa_{T}/\Jbar{L}$.  Note this depends on the applied pressure, via the mean box size $\overline{L}$.

We simulate high density systems where the typical free space per particle is much less than $ l_{0}$ and the mean collision time is much less than $\tauB$. 
In constant-density simulations we use $\phi=0.88$; for constant-pressure we take the corresponding value $P=7.33l_0^{-1}$. 
This gives a mean free space per particle of $0.136l_{0}$ and a typical collision time of $0.018\tauB$. 

\subsection{Large Deviations}

As anticipated in the introduction, we will be concerned here with ensembles of trajectories which are biased to non-typical values of an `activity' parameter $K$.  Our analysis follows that of \cite{Hedges:2009hb}.  A trajectory of the system is a realisation of the system developing through time. A trajectory has a time duration $\tobs$ which is composed of $M$ smaller and consecutive ``slices" of time, each of length $\Delta t$ such that $\tobs=M\Delta t$.

\begin{figure}
\includegraphics[width=8cm]{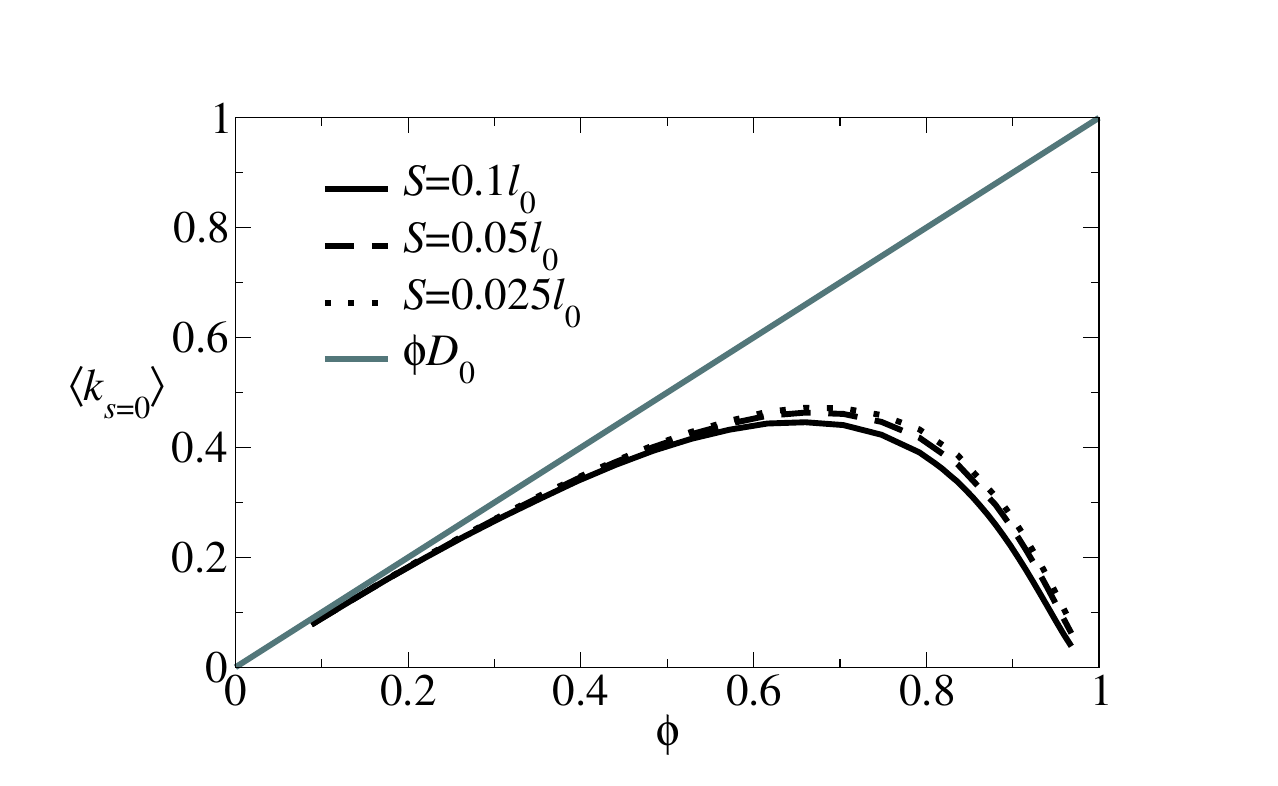} 
\caption{Equilibrium activity $k$ as a function of the packing fraction, $\phi$, 
and the maximum step size, $S$, in the constant-density ensemble.  The activity is linear in $\phi$ for small packing fractions, 
because $k$ is proportional to the number of particles per unit volume.  For larger $\phi$, particles start to obstruct
each other and the activity falls.
}
\label{fig:NVT_K0_rho}
\end{figure}

To define the activity $K$, we measure the squared displacements of all particles during each slice and sum over the $M$ slices in a trajectory:
\begin{align}
K\left[x(t)\right]=&\sum_{j=1}^{M} \sum_{i=1}^{N} \left| x_{i}\left(j\Delta t\right) - x_{i}\left((j-1)\Delta t\right) - \Delta\Jbar{x}_j\right|^{2} \label{eqn:defnK}
\end{align}
where the notation $[x(t)]$ indicates a functional dependence on all particle positions throughout a trajectory.
The quantity $\Delta\Jbar{x}_j=(1/N) \sum_i [ x_{i}\left(j\Delta t\right) - x_{i}\left((j-1)\Delta t\right) ]$ is the change of the centre of mass
during the $j$th slice: for large systems this has a negligible effect on $K$ but subtracting it in this way helps to minimize finite-size effects
in simulations.
It is often useful to normalise the activity as
\begin{align}
k=&\frac{K}{L\tobs} \label{eqn:defnk} .
\end{align}
(For the constant pressure systems, we replace $L$ by $\Jbar{L}$ in this equation.)
We take the time interval  $\Delta t=\tau_{B}$, so that $K$ depends on movement of particles on length scales comparable with their size, consistent with previous studies \cite{Hedges:2009hb}.
Fig.~\ref{fig:NVT_K0_rho} shows the dependence of the activity $k$ on the packing fraction $\phi$ and the MC step size $S$. 

The large deviation formalism uses a biasing field $s$ to investigate trajectories where $\tobs$ is large but
the activity $k$ deviates significantly from its typical value.  To this end, we define a probability distribution over the trajectories
of the system: 
\begin{align}
P_{s}[x(t)] & =\frac{P_{0}[x(t)]\myexp^{-sK[x(t)]}}{Z(s,\tobs)} \label{eqn:PsP0}
\end{align}
where $Z(s,\tobs)=\langle {\rm e}^{-sK}\rangle_{\rm eq}$ is a dynamical partition function (the average is evaluated at equilbrium).  
By analogy with equilibrium thermodynamics, we interpret
$\psi(s,\tobs) = -(L\tobs)^{-1} \log Z(s,\tobs)$ as a dynamical free energy (or free energy density).  For large $N$ and $\tobs$,
the free energy may develop a singular dependence on $s$, which signals the presence of phase transitions.

We use a notation $\langle \cdot \rangle_s$ to indicate averages with respect to the distribution (\ref{eqn:PsP0}).  
It is easily verified
 that the mean activity $k(s,\tobs) \equiv (L\tobs)^{-1} \langle K \rangle_s$ is obtained by a derivative of the free energy: $k(s)=(\partial \psi/\partial s)$, while the susceptibility
$\chi(s,\tobs) \equiv (L\tobs)^{-1} \langle (K - \langle K\rangle_s)^2 \rangle_s$ is equal to $-(\partial k/\partial s)$.

\subsection{Transition Path Sampling}

To sample the biased ensemble of trajectories given in (\ref{eqn:PsP0}), we use transition path sampling\cite{vanErp:2005ju,Bolhuis:2002ew}.   
An initial trajectory is generated from an equilibrium initial configuration by simulating the dynamics of a system for a duration $\tobs$.
A new trajectory is created by copying a randomly selected portion of the first trajectory to either the start or end of the new trajectory.
The rest of the new trajectory is then generated according to the relevant Langevin equations.  
The new trajectory is compared to the first and replaces it with probability:

\begin{align}
P_{\mathrm{accept}}=\mathrm{min}\left\{1,\myexp^{-s\Delta K}\right\}
\end{align}
where $\Delta K=K_{\mathrm{new}}-K_{\mathrm{old}}$.
Generation then continues using the most recently accepted trajectory as the parent. 
Using this method means that, after many iterations, the algorithm samples trajectories according to the distribution defined in~(\ref{eqn:PsP0}).

\section{Results -- constant-density}\label{sec:NVT_results}

In this section we present results for biased ensembles of this hard-particle system in the $NVT$ (constant density) ensemble.
All results are for the case $\phi=0.88$, as in~\cite{Jack:2015cx}.  
As discussed in that work, we expect many aspects of the system's behaviour to be independent of $\phi$.

\subsection{Phase separation for $s>0$}

\begin{figure}
\includegraphics[width=8cm]{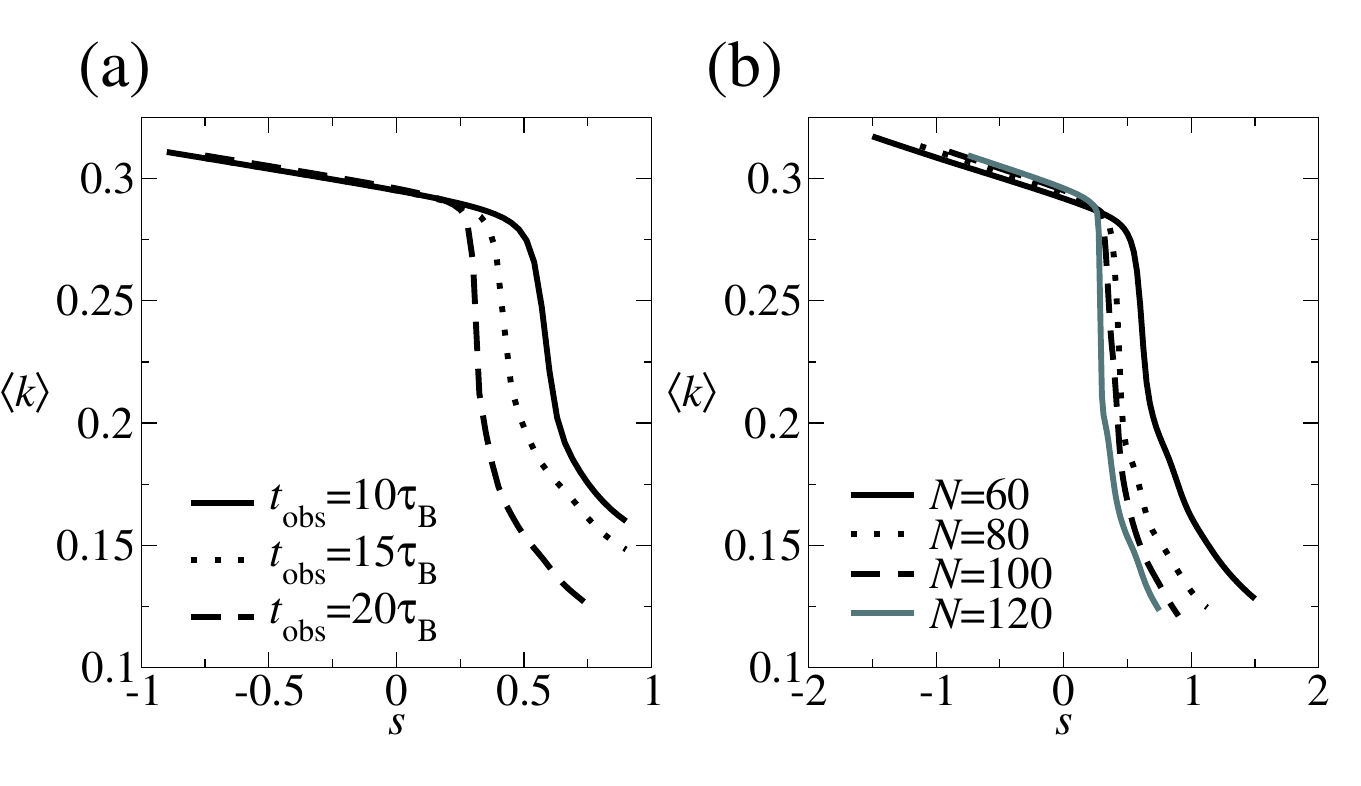}
\caption{The average intensive activity for constant-volume systems at $\phi=0.88$. (a) Behavior of the activity for a system of $N=100$ particles, varying $\tobs$. (b) The effect of changing the number of particles in the system at constant $\tobs=20\tauB$.}
\label{fig:KsRV}
\end{figure}

{We first consider the effect of a bias $s>0$, which leads to a phase transition in this system\cite{AppertRolland:2008coa,Lecomte:2012en,Jack:2015cx}.
Phase transitions are signalled by singularities in the free energy $\psi(s)$, which appear only in the limit when both the observation time $\tobs$ and the system size $N$ are very large. 
Fig.~\ref{fig:KsRV} shows the average activity $k(s)$ for different system sizes and observation times. 
In particular, Fig.~\ref{fig:KsRV}(a) shows the effect of increasing $\tobs$ at fixed system size $N=100$, while Fig.~\ref{fig:KsRV}(b) shows dependence on system size $N$, all obtained for large $\tobs=20\tauB$.
As in glass-forming systems, one observes a crossover at some $s=s^*$, from active to inactive dynamics; the value of $s^{*}$ is positive and depends on both $N$ and $\tobs$.

We note that for $s\to\infty$, the system must arrive at the state with minimal propensity for activity, which should be the fully phase-separated state, where the particles in the dense cluster are all touching each other.  
It is therefore clear that phase separation must occur at some field $s^*$.   
In fact, this phase transition can be predicted and analysed in detail in the framework of fluctuating hydrodynamics~\cite{Spohn:1983dt,Eyink:1990bx,bertini-rev}, which predicts that the system will phase separate whenever $(\partial^2/\partial\rho^2)\langle K \rangle_{\rm eq} <0$, and that $s^* \sim N^{-2}$ tends to zero as the system gets large\cite{AppertRolland:2008coa,Lecomte:2012en,Jack:2015cx}.  

\begin{figure}
\includegraphics[width=8.5cm]{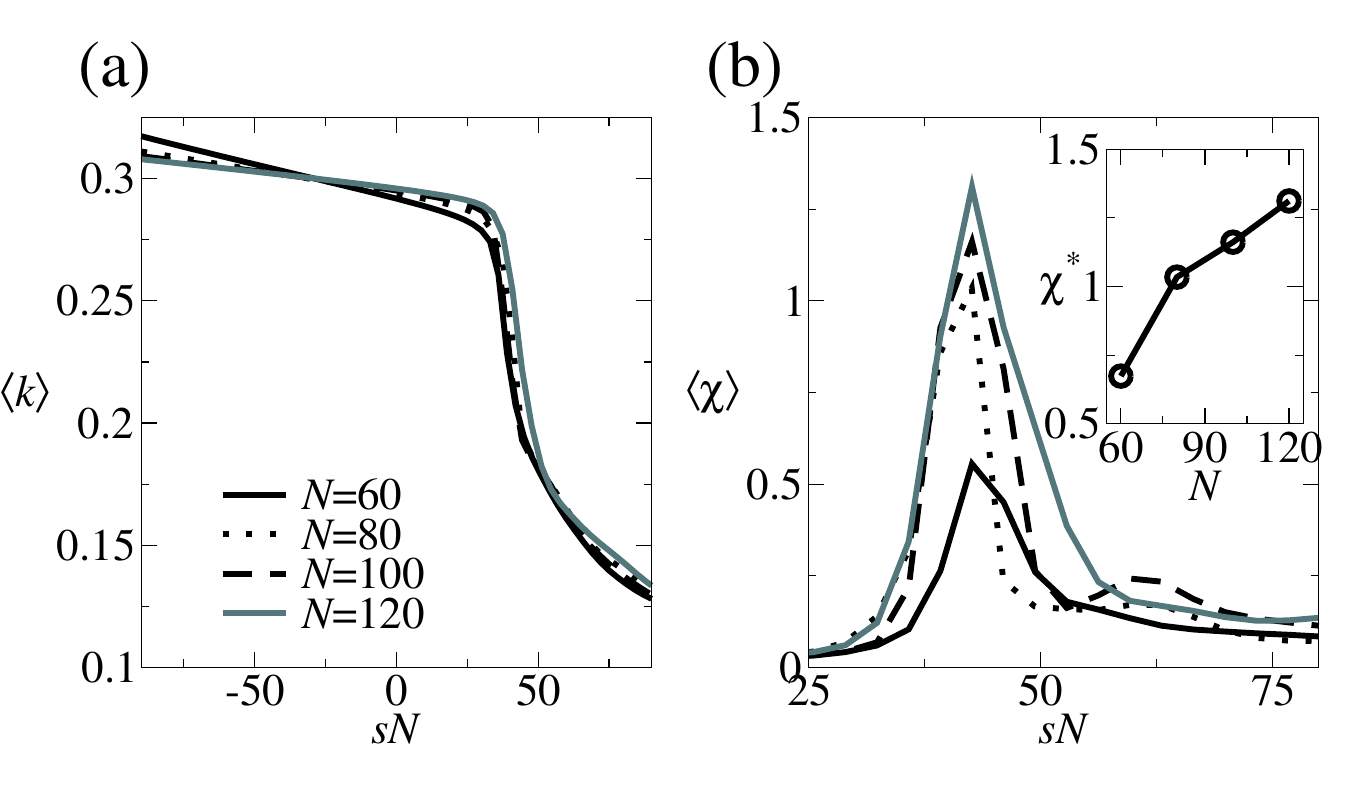}
\caption{Scaling of the transition with system size. (a) $k(s)$ collapses onto a single curve when scaled by $N$. (b) The peak in the dynamic susceptibility occurs at a constant-value of $s^{*}N$. Inset: The height of $\chi^{*}$ increases with increasing system size as the magnitude of fluctuations in the activity increases with $N$.}
\label{fig:KsRV_scaledbysN}
\end{figure}

Fig.~\ref{fig:KsRV_scaledbysN} shows a finite-size scaling analysis of the transition.  
For fixed $\tobs=20\tau_{\rm B}$ and increasing $N$, we find $s^* \sim N^{-1}$, as indicated by the collapse of $k(s)$ and the scaling of the peak in the susceptibility $\chi$. 
It seems that the theoretical prediction that $s^* \sim N^{-2}$~\cite{AppertRolland:2008coa,Lecomte:2012en} may be observable only for larger-$N$ and/or larger $\tobs$: we discuss this possibility in Section~\ref{subsec:stability-phasesep}.

\begin{figure} 
\includegraphics[width=6cm]{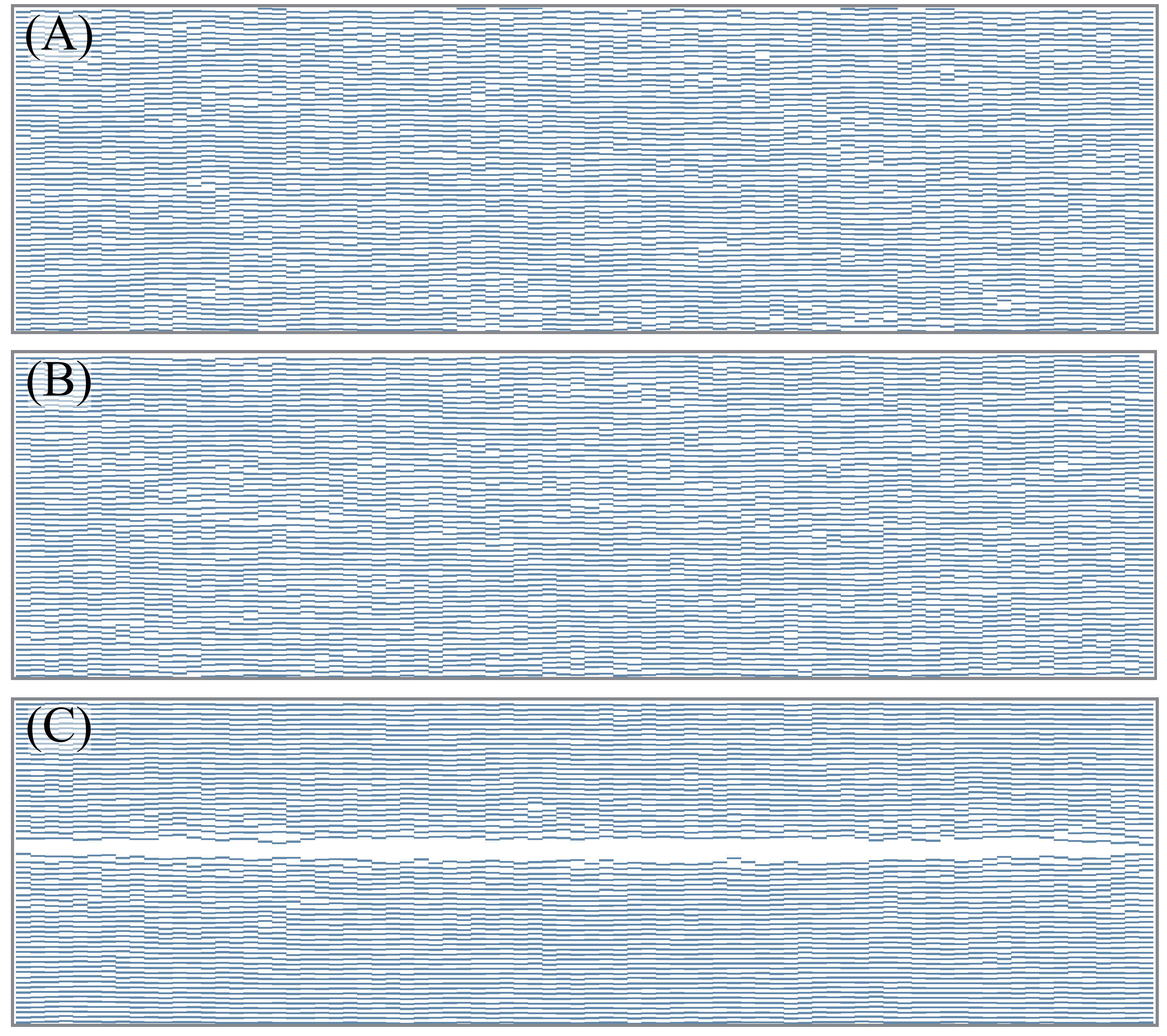}
\caption{Trajectories of a constant-density system at $N=60$, $\phi=0.88$ and $\tobs=20\tauB$. Blue boxes represent particles. Time runs horizontally, position vertically. Applied bias of (A) $s=-1$, (B) $s=0$, (C) $s=1.25$.}
\label{fig:RV_snapshots}
\end{figure}

\subsection{Structure of the phase separated state}

Fig.~\ref{fig:RV_snapshots} shows trajectories for the biased ensemble at several values of $s$, representing active, equilibrium and inactive trajectories. 
At $s=0$ the system is an equilibrium fluid of hard particles:
``bubbles'' of local free volume are seen throughout the trajectory, distributed randomly in time and space. 
For $s<0$ the system has slightly higher than equilibrium activity and exhibits no obvious structural change from equilibrium (see however Fig.~\ref{fig:NVT_rods_structure} below). 
In the inactive phase ($s>s^{*}$, Fig~\ref{fig:RV_snapshots}c) the system has fully phase separated for the whole trajectory.  Temporal boundary effects can be seen at the beginning and end of the trajectory where the phase separation deteriorates. 

\begin{figure}
\includegraphics[width=7cm]{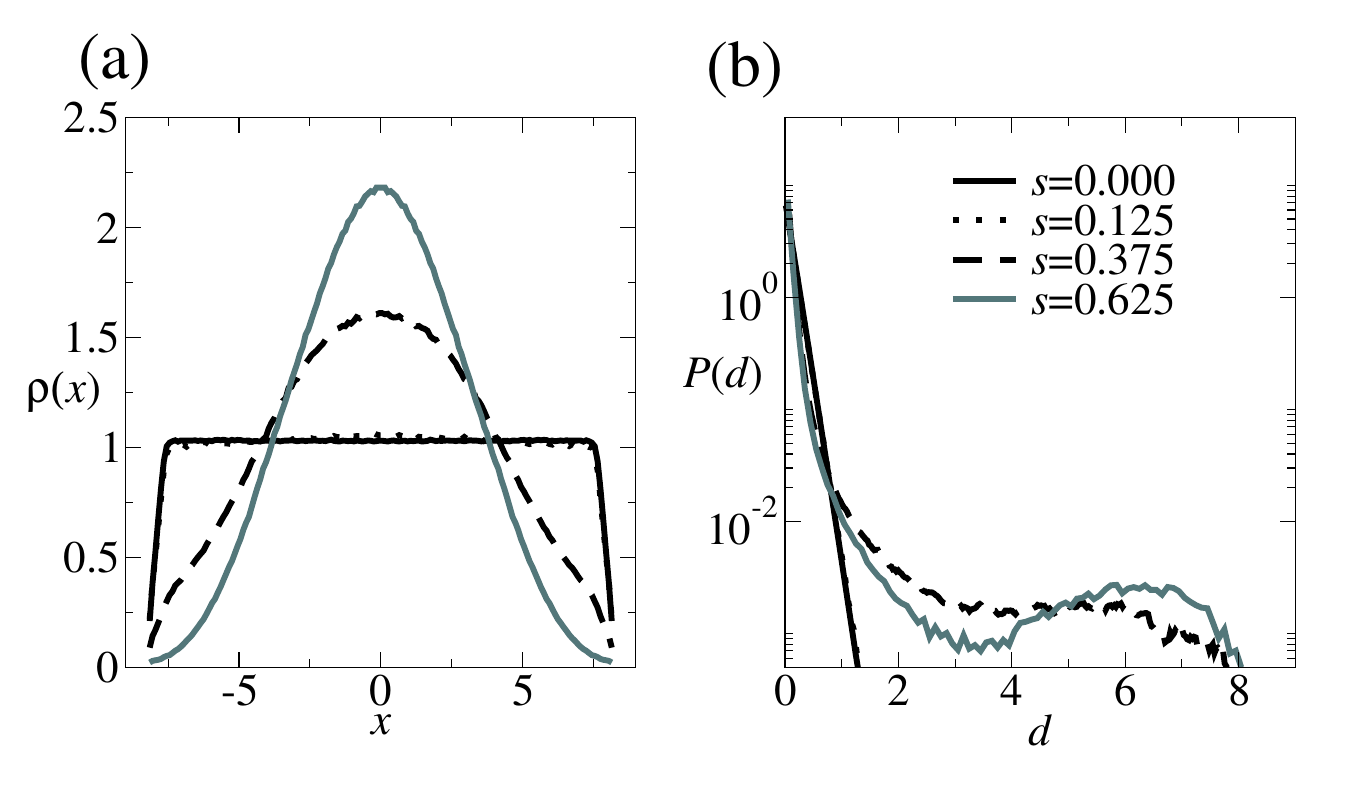}
\caption{Structural measurements in the constant-density regime with point particles at $\frac{N}{L'}=7.33$ with $N=120$ for equilibrium and inactive systems. (a) Measurements of the one-body density $\rho(x)$ for homogeneous and phase-separated states. (b) distribution of separations for the equilibrium and inactive phases. (The dotted lines $s=0.125$ are very close to the $s=0$ results.)}
\label{fig:NVT_Rods_phase_sep}
\end{figure}


To probe the structure of these systems it is convenient to make a change of co-ordinates. The system is one-dimensional and the particles are hard, so the ordering of the particle co-ordinates is fixed: there is no ``overtaking''.   We number particles so that their co-ordinates are in an increasing sequence, and define new co-ordinates $X_j=x_j-jl_0$ which are also ordered in the same way.
The $X_j$ are co-ordinates of point particles in a system of size $L'=L-Nl_0$.  At equilibrium, the positions $X_j$ are uncorrelated -- they represent positions of ideal gas particles.  For example, if we define a Fourier-transformed density $\delta\rho_q = \sum_j {\rm e}^{-iqX_j}$ and calculate the structure factor
\begin{equation}
S(q) = \frac{1}{L'} \langle \delta\rho_q \delta\rho_{-q} \rangle
\label{equ:sq}
\end{equation}
then we find $S(q) = N/L' = \frac{\phi}{l_{0}\left(1-\phi\right)}$, independent of $q$.

Now define $d_i$ as the separation between particle $i$ and its right neighbour,
\begin{align}
d_{i}=x_{i+1}-x_{i}  .
\end{align}
In order to investigate the one-body density profile associated with the phase-separated state illustrated in Fig.~\ref{fig:RV_snapshots}(c), it is necessary to fix an origin. We accomplish this by finding the largest `gap' $d_{i}$ in any configuration.  We choose a random point within that gap,
and we place the origin the maximal possible distance, $L'/2$, from that point.  Thus the origin almost certainly lies within the dense phase.  
The density of point-particles is then $\rho(X) = \sum_j \delta( X - X_j )$, where $X_j$ is now measured with respect to this new origin. 
We average $\rho(X)$ to obtain the one-body densities shown in Fig.~\ref{fig:NVT_Rods_phase_sep}(a).
At equilibrium, the density profile is uniform, as expected (up to weak boundary effects that arise because the origin was constrained
to lie in the largest gap).
As $s$ is increased, the phase separation in Fig.~\ref{fig:RV_snapshots} appears as a non-uniform density profile:
a large peak appears in $\rho(x)$, which corresponds to the clustering of particles near the origin.  

The distribution of separations, $P(d)$, was also recorded for systems at $s=0$ and $s>s^{*}$.  This is shown in 
Fig.~\ref{fig:NVT_Rods_phase_sep}(b). 
At equilibrium, one finds an exponential distribution, typical of a 1$d$ equilibrium fluid.  
However the distribution of separations in the inactive phase is bimodal.  
For small $d$, the distribution is approximately exponential but with a smaller characteristic length scale than the equilibrium fluid: this corresponds to particles within the dense region of the system. 
For larger $d$, there is a broad distribution of separations that comes from pairs of particles located on opposite sides of the large void (each configuration
contributes only a single sample to the large-$d$ peak, so the width of this peak appears only after averaging many configurations).
These separations are comparable to the system size and are indicative of phase separation.

\subsection{Stability of the phase separated state}
\label{subsec:stability-phasesep}

\newcommand{\ee}{{\rm e}}

The phenomenon of phase separation is not expected in one-dimensional systems at equilibrium (assuming that forces are short-ranged).
We therefore explore the physics behind this effect more detail. 
One can estimate the probability that the trajectory shown in Fig.~\ref{fig:RV_snapshots}(c) would occur at equilibrium, as follows.  
The particles within the large cluster are constrained by their neighbours and their contributions to the activity $K$ are necessarily small.  
The particles at the boundaries of the cluster cannot move into the cluster, but they may move away from it.  
The probability  that one of these particles nevertheless remains close to the edge of the cluster for the whole time $\tobs$ scales as ${\rm e}^{ - \gamma \tobs}$, where $\gamma$ is a parameter with units of time, proportional to the rate for diffusion of the particle away from the cluster.

The key point is that maintaining the integrity of the cluster requires only that the two particles at its boundaries do not move away.  
Hence, for large $\tobs$, the probability  $P_{\rm ps}$ of a phase-separated state at equilibrium should satisfy $\ln (P_{\rm ps}/P_{\rm eq}) \gtrsim -2\gamma \tobs$, where $P_{\rm eq}$ is the probability of a typical equilibrium trajectory.  
Within the $s$-ensemble, the ratio $  (P_{\rm ps}/P_{\rm eq}) $ is multiplied by a term $\ee^{s\Delta K}$, where $\Delta K$ is the difference in activity between equilibrium and phase separated states. 
Since the activity at equilibrium is extensive (that is, $\Delta K \sim \delta k N \tobs$), one therefore expects phase-separated states to dominate for $s \gtrsim 2\gamma/(N \delta k)$.  

This argument essentially reproduces the prediction of \cite{Jack:2006gda} for phase transitions in kinetically constrained models; see also~\cite{Garrahan:2007gw}. 
Assuming that $\gamma$ is independent of $N$, we predict $s^* \lesssim N^{-1}$, consistent with Fig.~\ref{fig:KsRV_scaledbysN}.  
As discussed above, the more refined analysis available from fluctuating hydrodynamics predicts $s^* \sim N^{-2}$~\cite{Lecomte:2012en}.  
That is, the bias required to stabilise phase separated states is even less than that predicted by the simple argument given here.  
The reason is that the interface between high- and low-density regions of the system may not consist of a single particle, but can be smoothed out: this acts to reduce $\gamma$ to a quantity of order $1/N$, further stabilising the phase separated state. 
However, it seems that the regime in which these smoothed out interfaces can be observed is not accessible within our simulations (where both $N$ and $\tobs$ are limited by the computational effort required.)

\subsection{Hyperuniformity for $s<0$}

\begin{figure}
\includegraphics[width=7cm]{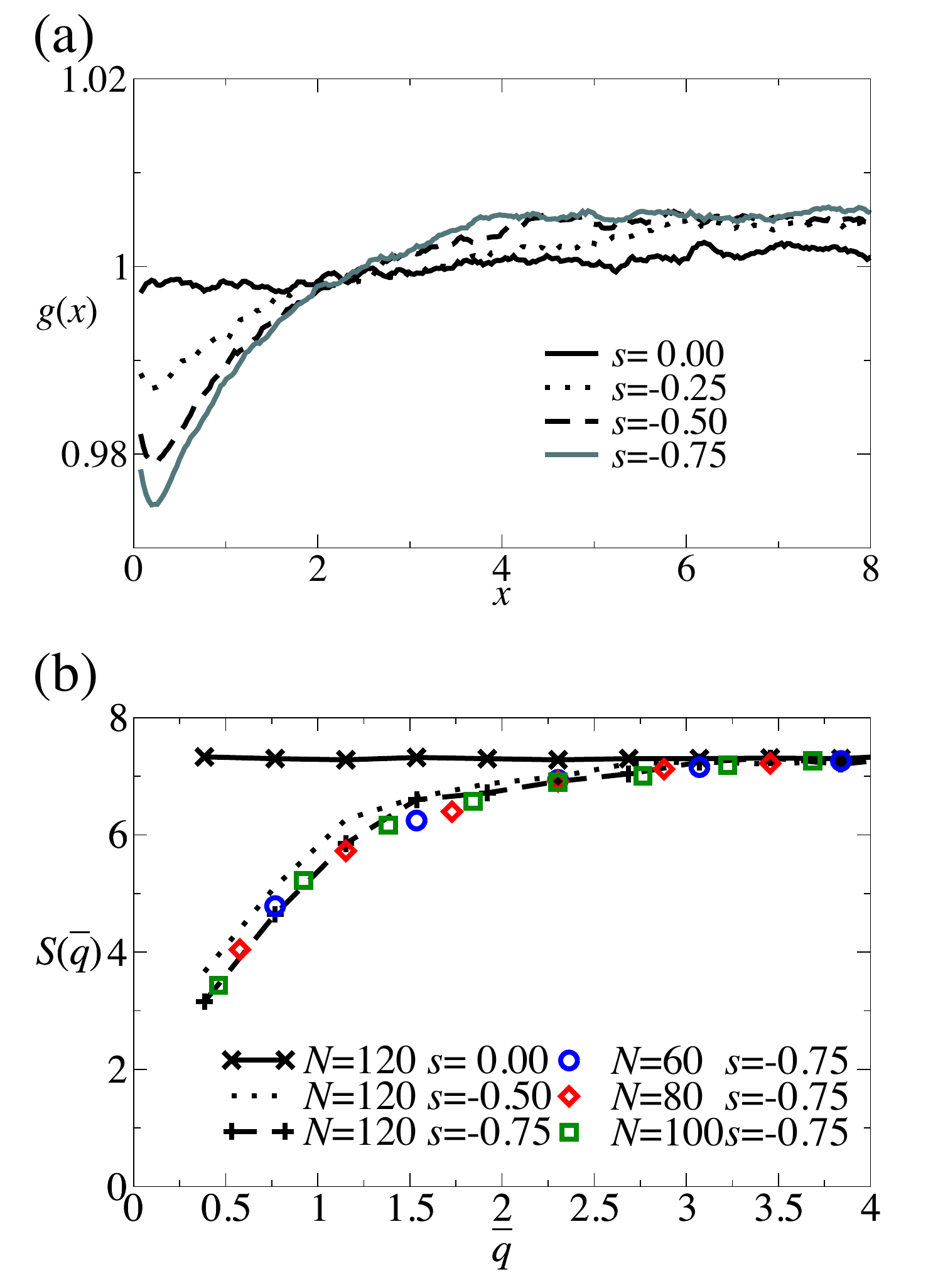}
\caption{(a) The pair correlation function, $g(x)$, for point particles in the constant-density regime with $N=120$, $\tobs =20\tauB$ in the active phase. The depression in $g(x)$ for small $x$ reflects an increase in local free volume around the particles as activity increases. (b) Small $q$ structure factor measurements for constant-density systems. Biasing to $s<0$ causes the suppression $S(q)\sim q$ consistent with the onset of hyperuniformity. At fixed $s$ all system sizes collapse onto a single curve.}
\label{fig:NVT_rods_structure}
\end{figure}

For $s<0$, the system is biased so that the particles move around more than they do at equilibrium.  
Fig.~\ref{fig:NVT_rods_structure}(a) shows the pair correlation function $g(x)=\langle \rho(x') \rho(x'+x) \rangle / \langle \rho \rangle^2$ in this regime.  
(The system is translationally invariant so there is no dependence on $x'$).
Particles appear to repel each other, leading to a depletion zone around each particle, which facilitates motion on small length scales.  However, the deviation of $g(x)$ from unity 
is small in absolute terms: this is a rather weak effect.

On the other hand, moving to Fourier space reveals a much stronger response to the bias $s$, associated with long-ranged correlations in the density.  This is illustrated in
Fig.~\ref{fig:NVT_rods_structure}(b), which shows the structure factor of the system for $s<0$. At small wavevectors (large length scales) the structure factor shows markedly different behaviour from the equilibrium fluid. At equilibrium the system behaves like an ordinary fluid with $S(q)=\mathrm{const.}$ for all $q$~\cite{Plischke:1492499}, recall (\ref{equ:sq}). Trajectories for $s<0$ have a very different structure factor as $q\rightarrow 0$. This suppression of long range density fluctuations is a sign of hyperuniformity\cite{Torquato:2003fk}: Hyperuniform states are distinguished by anomalously small density fluctuations on long length scales, eventually vanishing at infinite range, hence $S(q)\rightarrow 0$ as $q\rightarrow 0$.
The phenomenon has been found in jammed sphere packings~\cite{Jiao:dvnakV2q,Berthier:2010bk} and in a range of other physical
systems~\cite{Florescu:2009ev,Jiao:2014cp,hexner,Jack:2015cx}.

In numerical simulations, the size of the system limits our ability to investigate the small-$q$ behaviour of $S(q)$: we can obtain data only down to a minimum wave vector $q_{\mathrm{min}}=\frac{2\pi}{L'}$.
As shown in Fig.~\ref{fig:NVT_rods_structure}, density fluctuations at $q_{\mathrm{min}}$ are increasingly suppressed as $s$ becomes more negative. 
On increasing the system size, $q_{\mathrm{min}}$ is reduced, and $S(q_{\mathrm{min}})$ becomes smaller (for any given $s$).
Fig.~\ref{fig:NVT_rods_structure}(b) shows that all values of $S(q)$ collapse onto one curve at fixed $s$. 
This suggests that for any $s<0$ then:
\begin{align}
\lim_{N\rightarrow\infty}S(q_{\mathrm{min}},s)=0  
\label{eqn:KsRV_HU}
\end{align}
while for $s=0$ one has $\lim_{N\rightarrow\infty}S(q_{\mathrm{min}},0)=\frac{\phi}{l_{0}\left(1-\phi\right)}$.  
This discontinuous response to the field $s$ at $s=0$ corresponds to a dynamical phase transition to a hyperuniform state\cite{Jack:2015cx}.

\section{Results -- constant-pressure}\label{sec:NPT_results}

\subsection{Inactive state, $s>0$}

\begin{figure}
\includegraphics[width=8cm]{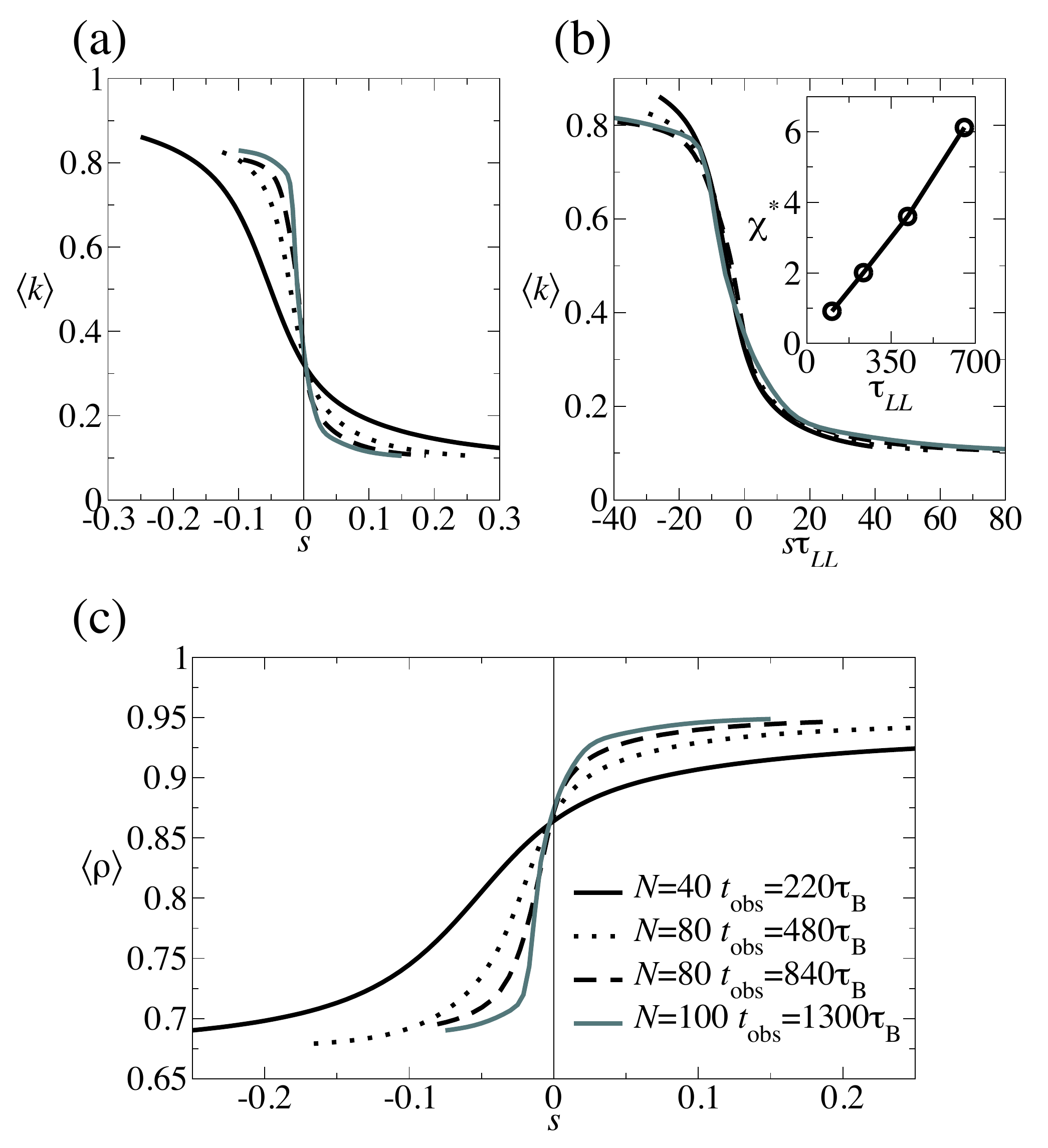}
\caption{Dynamic behaviour of the constant-pressure system.  All results come from 
trajectories of duration $\tobs\approx 2\tau_{LL}$ at $P=7.33$. (a) the intensive activity and dynamic susceptibility of the system as a function of $s$.
The transition (crossover) takes place at $s^{*}=0$ for all system sizes, and the width of the crossover is proportional to
 $(\tau_{LL})^{-1}$. (b) The data for all system sizes collapses when scaled by $\tau_{LL}$, (inset) the peak in the dynamic susceptibility scales with $\tau_{LL}$. (c) The (total) density of the system changes
as the system undergoes the transition shown in~(a).}
\label{fig:KsRP}
\end{figure}

\begin{figure}
	\includegraphics[width=8cm]{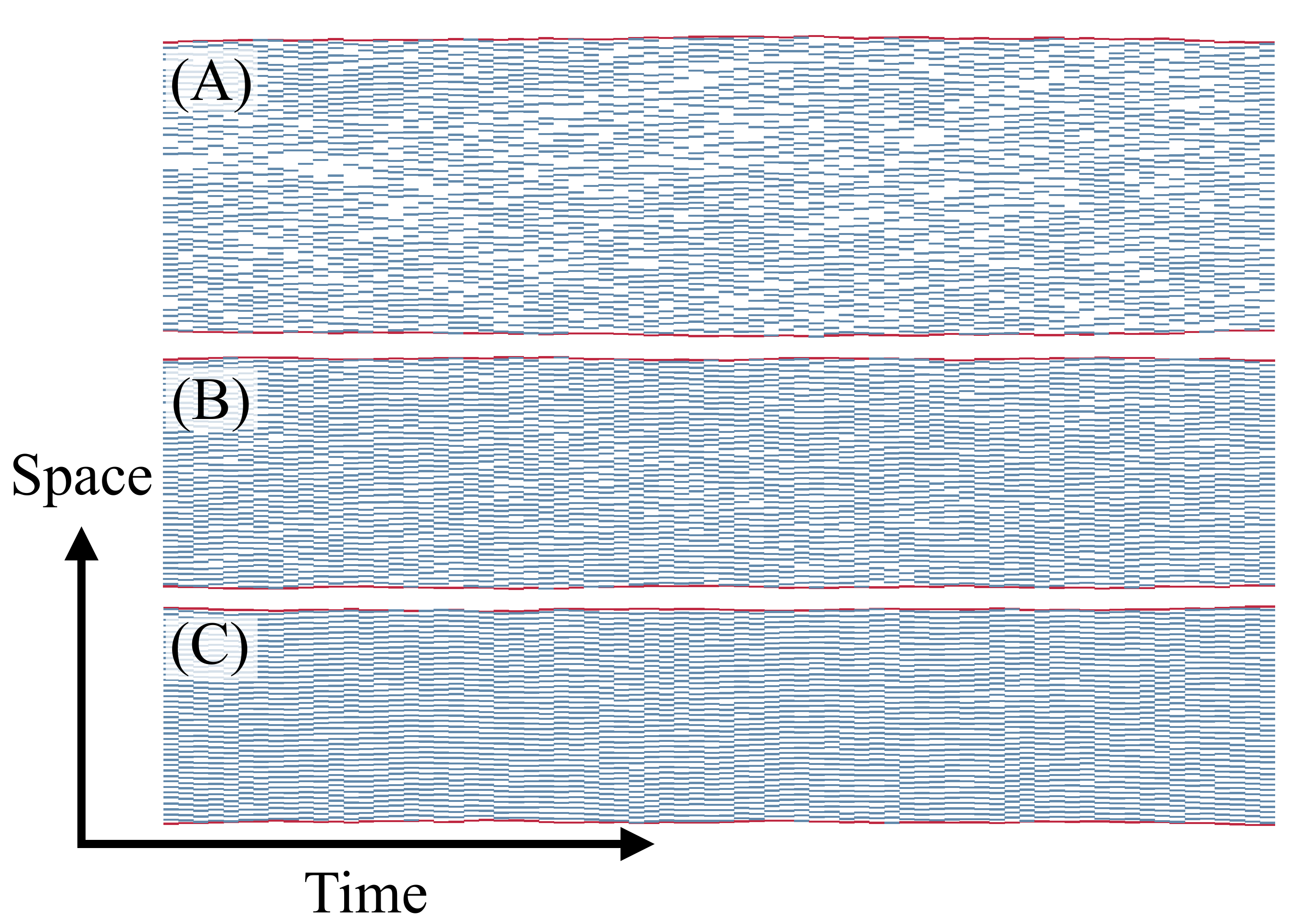}
	\caption{Representative trajectories of constant pressure systems at $P=7.33l_{0}^{-1}$ with $N=40$, $\tobs=220\tauB$ at different biases. Red boxes represent the boundaries of the system. (A) $s=-0.250$, the active systems increase in length and thus reduce global density. (B) $s=0$, the equilibrium system has a fluctuating volume but maintains $\left<\phi\right>=0.88$. (C) $s=0.375$, the inactive system is compressed relative to equilibrium and thus has suppressed activity.}
	\label{fig:NPT_snapshots}
\end{figure}

We now consider the constant-pressure version of the model, in which the system size evolves in time 
according to equation~(\ref{eqn:volume_langevin}).  
In equilibrium, one expects local properties of single phases to be independent of ensemble. 
For example, since the pressure is constant throughout an equilibrium system, one may think of the constant-pressure simulation as representing a subsystem of a very large constant-volume system.
One might expect the same equivalence to hold in biased ensembles at $s\neq0$, but we will see that the applied bias $s$ leads to a breakdown of ensemble-independence.  

We take the pressure $P=7.33 l_0^{-1}$ so that the mean volume fraction at equilibrium is $\langle\phi\rangle=0.88$, consistent with Sec.~\ref{sec:NVT_results}.  The effect of the biasing field $s$ is shown in Fig.~\ref{fig:KsRP}.  Comparing with Fig.~\ref{fig:KsRV}, a similar transition is apparent, but instead of a crossover at $s^*>0$ that depends on $N,\tobs$, one instead observes a crossover very close to the equilibrium point $s=0$.  Note that the values of $\tobs$ used here are significantly larger than those used in the constant-volume system: they are comparable with the volume relaxation time of the barostat $\tau_{LL}\sim \Jbar{L}^2$ (recall Section~\ref{subsec:MC_dynamics}). 
Fig.~\ref{fig:NPT_snapshots} shows representative trajectories from biased ensembles in the constant-pressure system. Comparing
with Fig.~\ref{fig:RV_snapshots}, no phase separation occurs.  We also note that the box size varies with $s$, consistent
with Fig.~\ref{fig:KsRP}(c).

\begin{figure}
	\includegraphics[width=8cm]{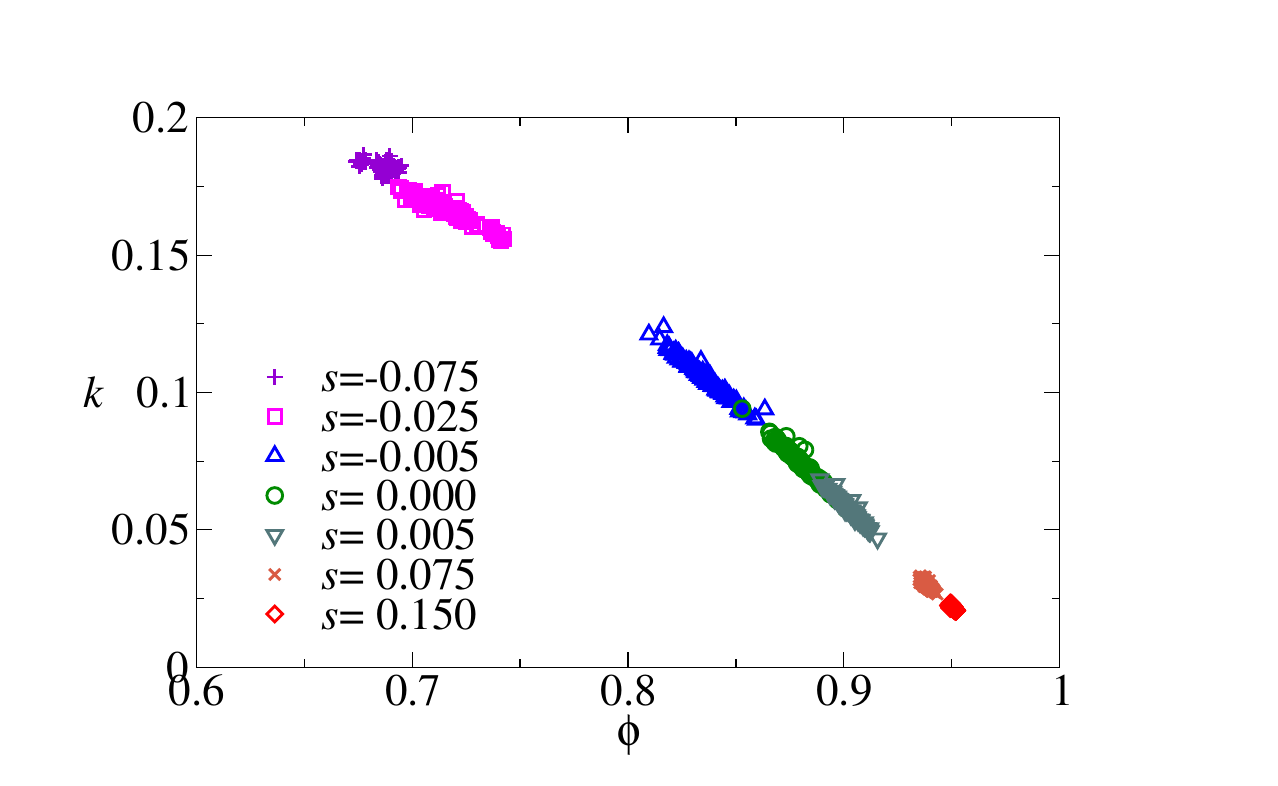}
	\caption{Scatter plot of activity against the average density of trajectories of duration $1300\tauB$ with $N=100$ particles from biased ensembles. There is a strong correlation between the activity and the density of the trajectories.}
	\label{fig:NPT_scatter_k_rho}
\end{figure}

Fig.~\ref{fig:NPT_scatter_k_rho} shows the correlation between activity and global density for all dynamic regimes.
Within fluctuating hydrodynamics, the density field is assumed to give a full description of the large-scale behaviour of this system,
In constant-volume systems, density fluctuations on finite wave vectors control the fluctuations in activity \cite{AppertRolland:2008coa,Jack:2015cx}. 
In the constant pressure system, Fig.~\ref{fig:NPT_scatter_k_rho} shows that the total density correlates strongly with the activity: high density is
associated with low activity, and vice versa.

Given this strong correlation, we can link the phase transition that takes place at $s=0$ in this system to the diverging hydrodynamic time scale $\tau_{LL}$
(recall Section~\ref{subsec:Models}).
We define the (normalised) autocorrelation function of the system size
\begin{equation}
C_{LL}(t) = \frac{ \langle \delta L(t') \delta L(t'+t) \rangle_{\rm eq} }{\langle \delta L(t')^2 \rangle_{\rm eq}  }
\end{equation}
which is evaluated at equilibrium (so there is no dependence on $t'$), with $\delta  L = L - \Jbar{L}$.  This correlation function decays on a time scale close to $\tau_{LL}$.
Similarly the correlation function of the activity is
\newcommand{\kk}{\mathcal{K}}
\begin{equation}
C_{kk}(t) = \frac{ \langle \delta \kk(t') \delta \kk(t'+t) \rangle_{\rm eq} }{\langle \delta {\kk}(t')^2 \rangle_{\rm eq}  }
\end{equation}
where $\kk(t)=\sum_i |r_i(t+\Delta t)-r_i(t)-\Delta \Jbar{x}(t)|^2$ is the quantity that appears in the definition of the activity $K$, recall (\ref{eqn:defnK}). 
To show the long-time behaviour of $C_{kk}(t)$ more clearly we smooth the function by convolving it with a Gaussian window, with variance $\sigma^{2}=\tauB^{2} /4$: we plot $\Gamma \sum_{t'}C_{kk}(t')\ee^{-2(t-t')^2/\tau_B^2}$, where the proportionality constant $\Gamma$ normalises the correlation function to unity at $t=0$. 

\begin{figure}
\includegraphics[width=8cm]{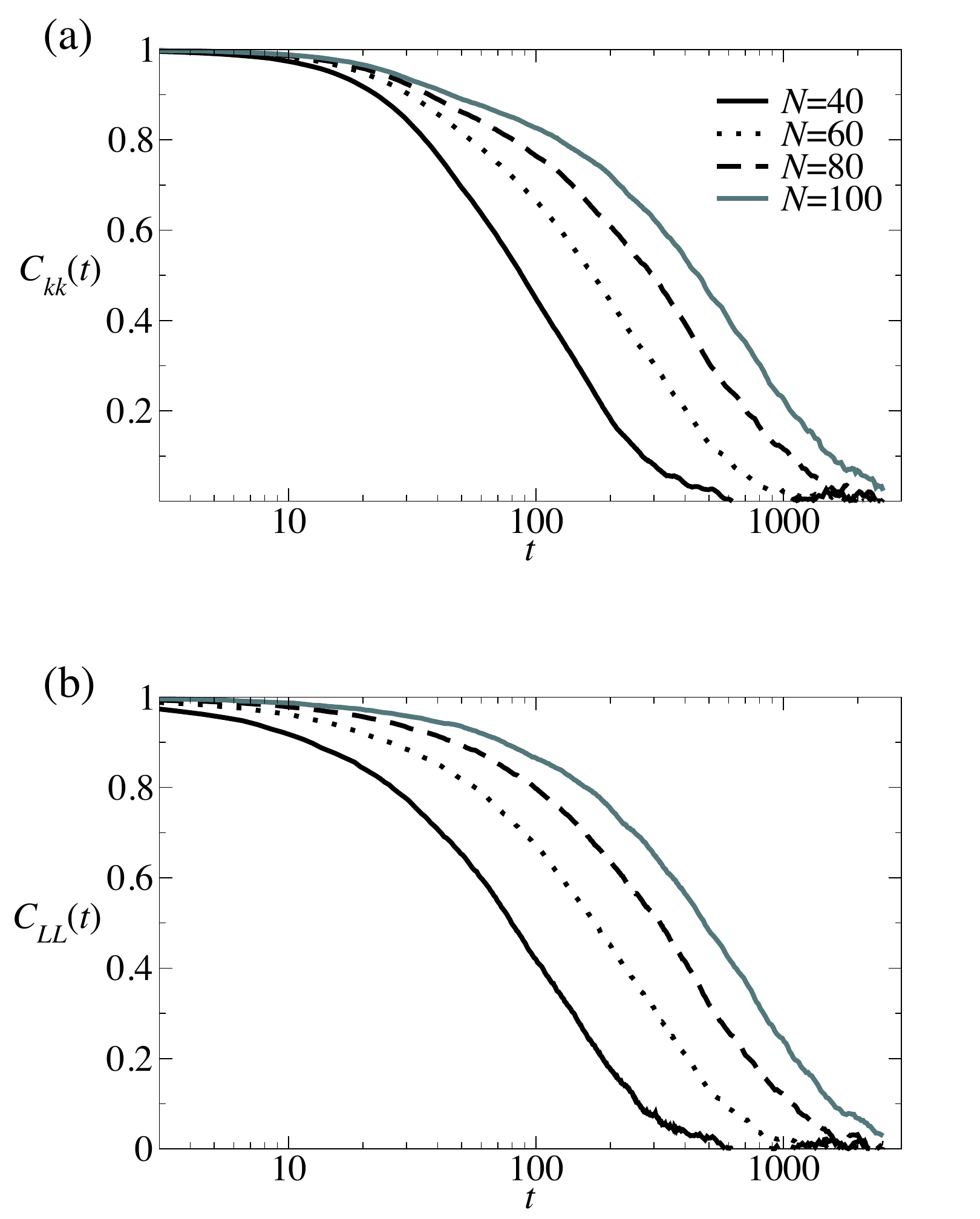}
\caption{The activity and volume correlation functions for  constant pressure systems at equilibrium, for different system sizes. The correlation time, $\tau_{kk}$, is strongly correlated with $\tau_{LL}$.  To display the long-time behavior most clearly, $C_{kk}(t)$ has been smoothed with a Gaussian window (see main text).}
\label{fig:NPT_rods_KKt_correl}	
\end{figure}

The correlation functions ${C}_{kk}(t)$ and $C_{LL}(t)$ behave very similarly, consistent with the idea that the activity fluctuations are strongly correlated with those of the global density (and hence to the system size). 
Since the volume relaxation time $\tau_{LL}$ diverges as $\Jbar{L}^2$, we therefore expect a similar divergence in the relaxation time of the activity.

This divergent time scale is important because the susceptibility $\chi$ is related to the autocorrelation function of the activity as
\begin{align}
\chi(s=0,\tobs\to\infty) &= \frac{2}{\Jbar{L}} \int_{0}^{\infty} \!\mathrm{d}t\, \langle \delta \kk(0) \delta \kk(t) \rangle_{\rm eq}
\label{equ:chi-tau}
\end{align}
so that $\chi\sim\tau_{kk}\sim \Jbar{L}^2$ diverges at $s=0$, which we interpret as a dynamical phase transition.  
(The equal time value of the correlator in this equation scales as $\Jbar{L}$ since $\delta \kk$ is extensive in the system size: this $\Jbar{L}$-dependence cancels with the prefactor so that the right hand side scales with $\tau_{kk}$, with a prefactor of order unity.)
Since $\chi=-dk(s)/ds$, the divergence of $\chi$ corresponds to a singularity in $k(s)$ and hence a dynamical phase transition.  
This amounts to a perturbative argument for the existence of the phase transition: a related perturbative argument based on fluctuations at finite wavevector was used in~\cite{Jack:2015cx} to explain the existence of phase transitions in systems at finite volume.

Recalling the discussion of Sec.~\ref{subsec:stability-phasesep}, the analogous argument for the constant-pressure system is that
the noise force $\eta_L$ in (\ref{eqn:volume_langevin}) acquires a finite average in the inactive state, resulting in a reduced system size.
The previous argument based on $\tau_{LL}$ indicates that biasing the noise force in this way requires very little cost in probability:
this low cost appears partly because only a single noise term needs to be biased, but also because the absolute size of the
bias becomes small in large systems, due to the scaling of the diffusion constant $D_L$ with system size (Eq.~(\ref{eqn:diffusion_relation})).

Finally we note that $s^*$ shifts from a value of order $1/N$ in the constant-density system to a value close to zero at constant pressure.  We interpret
the small positive $s^*$ (for the constant-volume system) in terms of the probability cost required to form the interface in a phase-separated system at constant density.  At constant pressure,
no interface is required so the system can have a diverging linear reponse, as shown in (\ref{equ:chi-tau}).

\begin{figure}
\includegraphics[width=7cm]{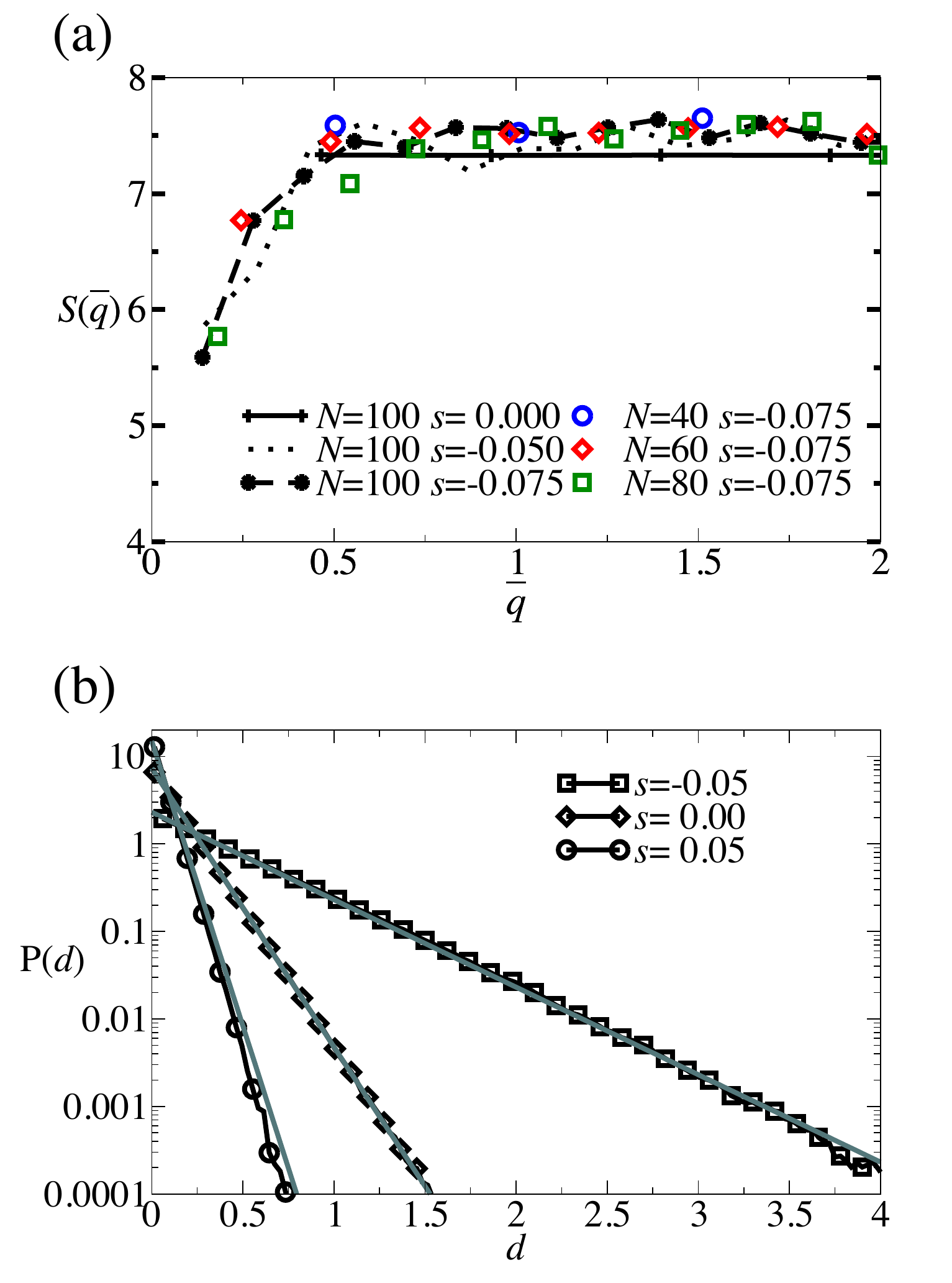}
\caption{(a) Structure factors in the constant-pressure regime when biased to higher than equilibrium activities. There is a suppression of long range density fluctuations as in the constant-density active phase. (b) The distribution of separations for a system of $N=100$ particles. Symbols represent measured distributions, solid lines are exponential distributions with a mean separation calculated from the mean volume. In all regimes the separations are distributed exponentially and are similar to equilibrium, albeit with a different mean separation.
}
\label{fig:NPT_rods_structure}
\end{figure}

\subsection{Active state, $s<0$}

In Fig.~\ref{fig:NPT_rods_structure}(a), we show the structure factor of the constant-pressure system for $s<0$.
For a given bias $s$, the fluctuations in the total system size are small in relative terms, so we evaluate the structure factor at wavevectors $q=2n\pi/L'$ as usual, and calculate $S(q)$ by an ensemble average at fixed $n$.  
This provides an estimate of $S(\Jbar{q})$ with $\Jbar{q}=2\pi n/\Jbar{L'}$.
The results of Fig.~\ref{fig:NPT_rods_structure}(a) are consistent with hyperuniformity of the active ($s<0$) phase, although the effect is weaker than that shown in Fig.~\ref{fig:NVT_rods_structure}, for the constant-density system.
We also show the distribution of particle separations in Fig.~\ref{fig:NPT_rods_structure}(b), for comparison with Fig.~\ref{fig:NVT_Rods_phase_sep}(b). 
The distribution fits well to an exponential form, independent of $s$.  
Given the correlations that are apparent from Fig.~\ref{fig:NPT_rods_structure}(a), this result is somewhat surprising: it might be that the correlations are sufficiently weak on short length scales that they are not discernible from $P(d)$.

In this constant-pressure system, it seems that achieving large deviations by changes in structure (for example phase separation or hyperuniformity) is unfavourable compared to changing the system density.  
Thus, particle separations remain exponentially distributed independent of $s$, but the system density depends strongly on $s$.
Even at the longest length scales and largest $|s|$ the structure is only weakly affected, compare Fig.~\ref{fig:NVT_rods_structure}(a) and Fig.~\ref{fig:NPT_rods_structure}(a).

\section{Discussion -- pressure balance and mechanical equilibrium}
\label{sec:virial}

To rationalise the numerical results shown here, it is useful to consider the balance of forces in the phase-separated
state shown in Fig~\ref{fig:RV_snapshots}c.  To this end, we consider the virial formula for the pressure.  As we briefly review here,
this expression can provide information on force balance even in systems that are not at equilibrium.  We define the virial by
\newcommand{\vir}{\mathcal{V}}
\begin{equation}
\vir = \sum_i x_i F_i
\label{equ:vir}
\end{equation}
where $F_i$ is the conservative (non-stochastic) force on particle $i$, which in our system is $F_i = v'(x_{i+1}-x_i) + v'(x_{i-1}-x_i)$, 
where $v(x)$ is the interparticle potential
(recall (\ref{equ:vx})) and the prime denotes a derivative.  (An alternative would be to include both conservative and stochastic forces
in the definition of $\vir$, but we choose this definition for later convenience.)

By restricting the sum in (\ref{equ:vir}) to particles within a given region of the system, and time-averaging 
over a period long enough for fast degrees of freedom to relax, we now show that we can obtain an estimate of the local
pressure.  Our discussion follows that of~\cite{hansen}.   We consider a region $B$ of the system, of linear size $L_B$.
For convenience, we assume that this region does not cross the periodic boundaries of the system.  
Then, if $\vir_B$ is the contribution to (\ref{equ:vir}) from particles $i$ in $B$ then 
\begin{equation}
\vir_B= \sum_{i\,\mathrm{in}\,B}  \sum_{j} x_i (\delta_{j,i+1} + \delta_{j,i-1} ) v'(x_j-x_i) .
\end{equation}
If we also decompose
the sum over $j$ into contributions from inside and outside $B$ then we obtain
\begin{equation}
\vir_B= \vir_B^{\rm int} + \vir_B^{\rm ext}
\label{equ:vie}
\end{equation}
with $\vir_B^{\rm int}=\frac12 \sum_{ij\,\mathrm{in}\, B} (x_j-x_i)v'(x_i-x_j) \delta_{j,i+1}$  and $\vir_B^{\rm ext} = \sum_{i\,\mathrm{in}\, B} x_i F^{\rm ext}_i$ where $F^{\rm ext}_i$ is the force
on particle $i$ from those particles outside $B$.  In a steady state, the forces on the two boundaries of the region $B$ must
balance.  Since the two boundaries are separated by a distance $L_B$ then $\overline{\vir_B^{\rm ext}}= -L_B P_B^{\rm mech}$, where the overbar
denotes a time average (over fast degrees of freedom) and $P_B^{\rm mech}$ is the average magnitude of $F^{\rm ext}$, which 
is the mechanical part of the pressure applied to the region $B$ by its environment.

In addition, recalling $\vir_B = \sum_{i\,\mathrm{in}\,B} x_i F_i$, the equation of motion (\ref{particle_langevin}) yields
\begin{align}
 \vir_B & = \frac{\Boltz T}{D_p}\sum_{i\,\mathrm{in}\,B} x_i ( \partial_t x_i - \eta_i ) \nonumber \\
& = \frac{\Boltz T}{D_p} \sum_{i\,\mathrm{in}\,B} [ \partial_t (x_i^2/2) - x_i\eta_i ]
 \end{align} 
For a system in a steady state, one has $\sum_{i\,\mathrm{in}\,B}\partial_t  (\overline{x_i^2})\approx 0$ and assuming that $\overline{\eta_i}=0$ then
$\overline{\vir_B}\approx0$.  In this case (\ref{equ:vie}) yields
\begin{equation}
P^{\rm mech}_B \approx \frac{\overline{\vir_B^{\rm int}} }{L_B}
\label{equ:pmechB}
\end{equation}
which is equivalent to the standard result for the mechanical part of the virial pressure.  
(In the usual case~\cite{hansen}, one considers Newtonian dynamics instead of overdamped Langevin dynamics, in which
the pressure includes an extra idea gas term $N\Boltz T/L_B$.  For overdamped dynamics, this term does not appear, but this has no
impact on mechanical equilibrium because the ideal gas contribution to the pressure is exactly constant throughout the system.)  
The approximate equality in (\ref{equ:pmechB}) appears because we have
invoked a time-average over fast degrees of freedom, instead of a full ensemble average. This distinction is useful when applying the
result to phase separated steady states such as that shown in Fig.~\ref{fig:RV_snapshots}c.

To consider those states,
the preceding argument must be generalised, because the biasing field $s$ can lead to noise forces $\eta_i$ with non-zero averages, $\overline{\eta}_i\neq 0$.  
In this case, balance of the total forces on region $B$ yields $\Delta P_B^{\rm mech} =  \frac{\Boltz T}{D_p}\sum_{i\,\mathrm{in}\,B} \overline{\eta_i}$ where
$\Delta P_B^{\rm mech}$ is the difference in mechanical pressure between the two sides of the region.  Considering a small region $B$
centred at position $x$, we obtain a force-balance
equation
\begin{equation}
\nabla P^{\rm mech}(x) = \rho \overline{\eta}(x)  \frac{\Boltz T}{D_p}
\label{equ:forcebal}
\end{equation}
where the noise force is averaged over particles in the vicinity of position $x$.  Again for small boxes (of size $\delta x$), one finds 
$\overline{\vir}(x) = - \frac{\Boltz T}{D_p} x\delta x\rho\overline{\eta}(x)$ and $\overline{\vir}^{\rm ext}(x) = -\delta x P^{\rm mech}(x) -x \delta x\nabla P^{\rm mech}(x)$.
Hence from (\ref{equ:vie},\ref{equ:forcebal}):
\begin{equation}
P^{\rm mech}(x) = \frac{\overline{\vir^{\rm int}(x)}}{\delta x} .
\end{equation}
That is, the virial still gives a useful estimate of the local pressure, even in the presence of noise forces with non-zero averages.  Note however that in the presence
of these noise forces, the mechanical pressure is not constant in space, but varies according to (\ref{equ:forcebal}).

Returning to the phase separated state in Fig.~\ref{fig:RV_snapshots}c, we see that the dense and sparse regions of the system are both
homogeneous, so we expect the local virial $\overline{\vir^{\rm int}(x)}$  and the mechanical pressure to be constant within each phase.  However, it is easily
verified that the virial differs strongly between the phases, so the mechanical pressure also differs.  The origin of this pressure difference
is the presence of non-zero noise forces at the boundaries of the cluster, consistent with (\ref{equ:forcebal}) and the discussion of 
Sec.~\ref{subsec:stability-phasesep}.
Given this observation, it is not surprising that phase coexistence was not observed in the constant pressure system for $s>0$: the 
phases that coexist in the contant-volume system do not have equal pressures.  

We note in passing that this argument may be straightforwardly generalised to 
 active matter systems~\cite{Wittkowski:2014dt,solon-press}, in which case the noise forces $\eta_i$ should be replaced by the particles' self-propulsion forces.  
In scalar active matter, one may find
coexistence between states with different mechanical pressures, due to the presence of one-body forces with non-zero averages.

\section{Conclusion}\label{sec:conclusion}

We have demonstrated dynamic phase transitions in a one-dimensional model of diffusing particles, including transitions from simple equilibrium fluid states into both
high-activity and low-activity states.  We considered both constant-density and constant-pressure systems: their transitions 
share some common features but there are also important differences. 
Based on the theory of fluctuating hydrodynamics, we argued~\cite{Jack:2015cx} that the transitions occur for all densities $\rho$, and the arguments given here indicate that this should also hold for all applied pressures.

Considering first the transition to inactive states, the constant-volume system undergoes phase separation, while the constant-pressure system
increases the local density.  (As in equilibrium systems, the constant-pressure system avoids interfaces between coexisting phases.)  
For large systems, the inactive phase occurs for all $s>0$, in both ensembles.  However, there is no signature of phase
coexistence in the constant-pressure system: we find only a dense phase, consistent with the different (mechanical) pressures of the 
coexisting phases in the contant-volume system.

For transitions to high-activity states, the constant-volume system spontaneously suppresses long range density fluctuations and develops a hyperuniform 
structure~\cite{Jack:2015cx}.  At constant pressure, the main feature of the high-activity state is that the total density decreases sharply, although there is
also some suppression of large-scale density fluctuations.  

Overall, these results emphasize that equilibrium ideas of ensemble equivalence do not apply directly when considering large
deviation phenomena such as those considered here.  While the mechanical pressure and the virial can still be related, the possibility
of phase coexistence at unequal pressures shows how familiar equilibrium concepts such as phase separation
need to be re-evaluated and generalised in these non-equilibrium settings.

\bigskip

\noindent
{\bf Acknowledgements.} We thank Peter Sollich for helpful discussions, and the
the EPRSC for support through grant EP/I003797/1.


\end{document}